\documentclass[10pt,conference]{IEEEtran}

\IEEEoverridecommandlockouts
\usepackage{cite}
\usepackage{amsmath,amssymb,amsfonts}
\usepackage{graphicx}
\usepackage{textcomp}
\usepackage{xcolor}
\usepackage{fancyhdr}
\usepackage{lipsum}
\usepackage{listings}
\usepackage{float}
\usepackage{makecell}  
\usepackage{tabularx}  
\usepackage{rotating} 
\usepackage{caption}
\renewcommand{\baselinestretch}{0.96}

\def\BibTeX{{\rm B\kern-.05em{\sc i\kern-.025em b}\kern-.08em
    T\kern-.1667em\lower.7ex\hbox{E}\kern-.125emX}}
   
\usepackage[ruled,linesnumbered]{algorithm2e}
\SetKwInput{Require}{Require}
\SetKwInput{Ensure}{Ensure}
\usepackage{amsmath}
\SetAlCapNameFnt{\footnotesize}
\SetAlCapFnt{\footnotesize}

\begin{document}
\title{VECA: Reliable and Confidential Resource Clustering for Volunteer Edge-Cloud Computing\thanks{This material is based upon work supported by the National Science Foundation (NSF) under Award Number: OAC-2232889. Any opinions, findings, and conclusions or recommendations expressed in this publication are those of the authors and do not necessarily reflect the views of the NSF.}}

\author{
\IEEEauthorblockN{
Hemanth Sai Yeddulapalli\IEEEauthorrefmark{1},
Mauro Lemus Alarcon\IEEEauthorrefmark{1},
Upasana Roy\IEEEauthorrefmark{1},
Roshan Lal Neupane\IEEEauthorrefmark{1}, \\
Durbek Gafurov\IEEEauthorrefmark{1},
Motahare Mounesan\IEEEauthorrefmark{2},
Saptarshi Debroy\IEEEauthorrefmark{2},
Prasad Calyam\IEEEauthorrefmark{1}}
\IEEEauthorblockA{
\IEEEauthorrefmark{1}University of Missouri-Columbia, USA; 
\IEEEauthorrefmark{2}City University of New York, USA\\
Email: 
\IEEEauthorrefmark{1}\{hygw7, lemusm, u.roy, neupaner, durbek.gafurov, calyamp\}@missouri.edu;\\
\IEEEauthorrefmark{2}mmounesan@gradcenter.cuny.edu; 
\IEEEauthorrefmark{2}saptarshi.debroy@hunter.cuny.edu
\thanks{
}
}
}

\maketitle

\begin{abstract}
Volunteer Edge-Cloud (VEC) computing has a significant potential  to support scientific workflows in user communities contributing volunteer edge nodes. However, managing heterogeneous and intermittent resources to support machine/deep learning (ML/DL) based workflows poses challenges in resource governance for reliability, and confidentiality for model/data privacy protection. There is a need for approaches to handle the volatility of volunteer edge node availability, and also to scale the confidential data-intensive workflow execution across a large number of VEC nodes. In this paper, we present VECA, a reliable and confidential VEC resource clustering solution featuring three-fold methods tailored for executing ML/DL-based scientific workflows on VEC resources. Firstly, a capacity-based clustering approach enhances system reliability and minimizes VEC node search latency. Secondly, a novel two-phase, globally distributed scheduling scheme optimizes job allocation based on node attributes and using time-series-based Recurrent Neural Networks. Lastly, the integration of confidential computing ensures privacy preservation of the scientific workflows, where model and data information are not shared with VEC resources providers. We evaluate VECA in a Function-as-a-Service (FaaS) cloud testbed that features OpenFaaS and MicroK8S to support two ML/DL-based scientific workflows viz., G2P-Deep (bioinformatics) and PAS-ML (health informatics). Results from tested experiments demonstrate that our proposed VECA approach outperforms state-of-the-art methods; especially VECA exhibits a two-fold reduction in VEC node search latency and over 20\% improvement in productivity rates following execution failures compared to the next best method.

\end{abstract}

There has been an exponential growth of data-intensive applications that rely on automation of complex scientific workflows with high demand for computational resources. To address this demand for processing power, the paradigm of volunteer computing~\cite{seti,vecflex} has evolved, offering a decentralized solution that harnesses the collective power of sporadically available edge compute resources (e.g., laptops, desktops, servers) contributed voluntarily by individuals or organizations~\cite{edgeenabler}. In recent times, the paradigm of Volunteer Edge-Cloud (VEC) computing has become possible due to the emergence of frameworks such as BOINC~\cite{BOINC} and Kubernetes-at-the-Edge~\cite{kubeedge}, where participants from scientific application communities unite to form a distributed computing ecosystem (such as e.g., Open Science Data Federation~\cite{osdf}) to benefit each other for tasks related to scientific data analytics using machine/deep learning~\cite{edgeenabler}. 

The inherent intermittent and volatile availability of VEC node resources can lead to unpredictable workflow performance~\cite{edgeenabler}, unlike the predicatable performance typically observed in data centers or cloud platforms with unconstrained/high-availability resources. This is followed by the resource management complexity in VEC environments, where VEC node providers can unexpectedly alter resource configurations, expose workflows to threat actors, and ultimately impact the capacity, trust, and availability of VEC node resources~\cite{fog-edge-resmngmnt-survey}. Consequently, for the orchestration of a potentially large scale of VEC nodes, the cloud hub needs to address challenges in scheduling the workflows with considerations to resource governance for reliability (to handle volatility in VEC node availability), and also confidentiality for model/data privacy protection (to avoid exposing model and data information to the VEC resource providers).

There is dearth of work that can address the scalability, reliability, privacy, governance, and communication overhead issues between nodes and job schedulers in VEC computing environments, posing the need for efficient mechanisms for managing a large number of diverse and unpredictable nodes~\cite{survey-iot-edge}. Recent works such as VECFlex~\cite{vecflex} and VELA~\cite{vela} offer solutions for scalable, and reliable services to manage the edge-cloud continuum, however they are not effective in handling the intermittent nature of VEC nodes and also in ensuring confidentiality of ML/DL-based scientific workflows. 

In this paper, we present a volunteer edge-cloud allocation (VECA) framework, which provides a reliable and confidential VEC resource clustering solution for executing ML/DL-based scientific workflows on VEC computing resources. The VECA solution features three methods to effectively cope with the intermittent nature of VEC nodes, and to ensure trusted computing with stringent data and model confidentiality. Firstly, a capacity-based clustering method using the $k$-means algorithm is presented to aggregate VEC nodes based on their capacity similarity (e.g., CPU, RAM, storage) thereby minimizing VEC node search latency of the scheduler. Secondly, a novel two-phase, globally distributed scheduling scheme is proposed to optimize job allocation based on workflow capacity specifications by using time-series forecasting using recurrent neural networks (RNN) and fail-over governance mechanism to improve productivity rate. Lastly, a confidential computing certifier is integrated to ensure privacy-preservation of the ML/DL-based scientific workflows within a trusted execution environment (TEE), using the AWS Nitro Enclaves~\cite{aws-nitro} as an exemplar implementation.

We evaluate VECA in a Function-as-a-Service (FaaS) emulation testbed that features OpenFaaS~\cite{OpenFaaS} and MicroK8S~\cite{MicroK8s} technologies, and uses the Amazon Web Services (AWS) cloud platform capabilities. We consider two ML/DL-based scientific workflows viz., G2P-Deep (bioinformatics)~\cite{g2pdeep} and PAS-ML (health informatics)~\cite{pasml} that are setup as Docker containers and are executed as serverless functions (i.e., without the need for provisioning specific servers) on scheduled VEC nodes. We implement the VECA components as microservices that use REST APIs and Message Queues for interaction. Our evaluation results demonstrate that our VECA approach significantly reduces VEC node search latency compared to existing baseline solutions i.e., VECFlex~\cite{vecflex}, and VELA~\cite{vela}. Furthermore, we evaluate how VECA enhances the overall productivity rate with availability prediction using RNN and fail-over capability using a governance strategy involves distributed cache management implemented via Redis\cite{Redis}.

The remainder of this paper is organized as follows: Section~\ref{sec:related_work} presents the related work. Section~\ref{sec:problem_description} describes the VECA problem formulation and outlines the proposed solution. Section~\ref{sec:methodology} details the $k$-means clustering approach. Section~\ref{sec:2-phasescheduler} discusses the two-phase scheduling and confidential computing based scientific workflow execution. Section~\ref{sec:experimentation_results} details the performance evaluation. Section~\ref{sec:conclusion_futureworks} concludes the paper and outlines future directions.

\vspace{-1mm}
\section{Related Work}
\label{sec:related_work}

\vspace{-1mm}
\subsection{Clustering for Volunteer Edge-Cloud Computing}
Recent works in the edge-cloud continuum such as VELA~\cite{vela} have set the precedent for distributed scheduling systems that effectively bridge the gap between cloud and edge computing realms. While this approach points out the inefficiencies of random cluster selection, it lacks specificity in considering the characteristics/behavior of the nodes for cluster selection. Similarly, CLARA~\cite{clara} highlights the advantages of leveraging clustering to enhance resource availability but fails to address the problem of efficient VEC node search and cluster-based resource allocation. Other recent works such as VECFlex~\cite{vecflex} and Greedy-Random~\cite{mounesan2024} address the brokering of VEC nodes for execution of data-intensive scientific workflows, however they do not consider clustering of the VEC nodes to meet workflow demands. The survey~\cite{survey-ml-scheduling-sla} provides details on how unsupervised learning algorithms such as $k$-means can be used for clustering the workloads. Inspired by the above works, VECA solution advances prior research by introducing an intelligent capacity-based clustering approach to reduce the search space in VEC computing for increasing efficiency in ML/DL-based scientific workflow scheduling.

\vspace{-1mm}
\subsection{Distributed Scheduling for Intermittent Availability}
The inherent complexities and the sporadic nature of volunteer resource contributions impact both the capacity and reliability of VEC systems, as noted by~\cite{edgeenabler, reliability-microblogging}. This issue is compounded by the need for sophisticated scheduling mechanisms capable of handling the unpredictable availability of resources. Prior works such as OneEdge~\cite{oneedge} and Mesos~\cite{mesos} underscore the importance of geo-distributed infrastructure and sophisticated resource sharing mechanisms, yet they do not address distributed scheduling challenges related to intermittent node availability. In contrast, approaches such as the application-aware task scheduling discussed in~\cite{application-aware-ts} partly address node volatility issues, however they do not align well with issues on volunteer resource dynamics. Addressing the unpredictable resource availability in a VEC environment, advancements have been made through stochastic models and semi-Markov processes as explored by~\cite{semi-markov-reliability, 8457994, 8473386}. These research studies provide a foundation for predictive analytics, which is crucial for anticipating resource availability and managing disruptions in VEC environments, as elaborated by~\cite{Tuli2020DynamicSF}. The findings in these works justify the predictive analytics component of our VECA approach that features a novel time-series based RNN model to address issues of distributed scheduling of VEC nodes with intermittent availability.

\begin{figure*}[t]
\centering
\includegraphics[width = 0.80\linewidth]{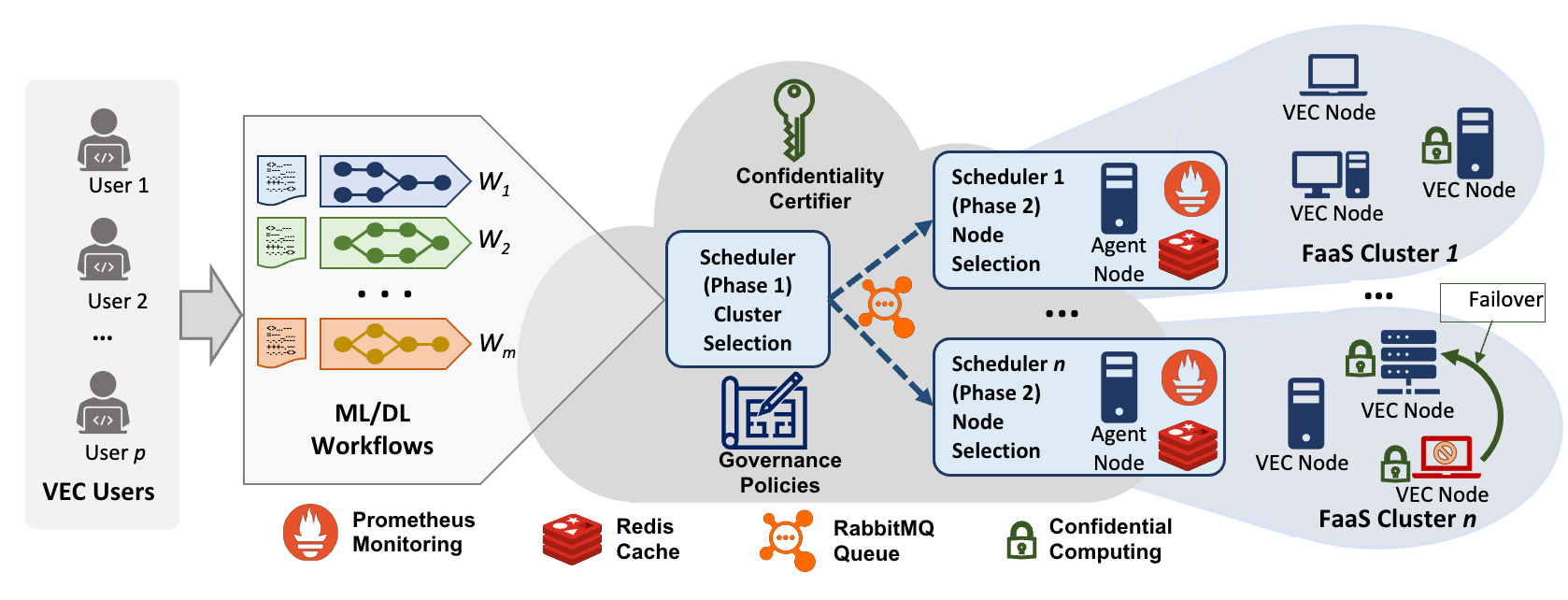}
\caption{\footnotesize{The VECA solution architecture illustrates users submitting ML/DL-based workflows to a Cloud Hub. Here, volunteer resources are clustered using the $k$-means algorithm and secured through a confidential computing framework. Two-phase distributed scheduling mechanism selects the most suitable cluster and the optimal VEC node within the selected cluster to execute the submitted workflow and meet user performance and security requirements.  
}
}
\label{fig:vec-solution}
\vspace{-6mm}
\end{figure*}

\section{Problem Formulation and Solution Overview}
\label{sec:problem_description}
In this section, we first discuss the problem in execution of scientific workflows via the management of dynamic volunteer resources, while addressing security and privacy concerns in VEC environments. Subsequently, we present our VECA solution overview to optimize VEC node resource allocation, manage the intermittent nature of volunteer resources, and preserve the privacy and confidentiality of data and models in ML/DL-based scientific workflow execution.

\vspace{-2mm}
\subsection{Executing Scientific Workflows in VEC Environments}

\subsubsection{Challenges}  
Within a VEC environment, we encounter tasks and resources with diverse needs and specifications. On one side, there are workflows with specific performance and security requirements, while on the other side, there exists a large number of volunteer resources with disparate specifications and security setups, as illustrated in Fig.~\ref{fig:vec-solution}. Current distributed scheduling systems include clustering approaches to group VEC nodes based on specific factors and manage the allocation of resources at the cluster level~\cite{vecflex, vela, clara}. However, these approaches lack effective mechanisms to manage the large number of volunteer resources in a manner that aligns with the dynamic and heterogeneous requirements of workflows. This gap results in suboptimal cluster selection, reducing the overall efficiency of resource allocation and utilization in VEC environments. The challenge lies in designing and implementing a clustering mechanism that can effectively map the computational requirements of workflows to the capabilities of available VEC node resources. The objective is to optimize the alignment between task demands and VEC node capabilities, minimizing resource allocation overhead and enhancing system responsiveness.

Given a set of VEC nodes \(N = \{n_1, n_2, \ldots, n_m\}\) and a set of ML/DL-based scientific workflows \(W = \{w_1, w_2, \ldots, w_k\}\), where each workflow \(w_j\) has defined resource requirements \(R_j = (r_{j1}, r_{j2}, \ldots, r_{jp})\) across \(p\) parameters (CPU, RAM, and Storage), the goal is to optimally cluster \(N\) nodes into \(k\) clusters \(C = \{c_1, c_2, \ldots, c_q\}\) such that:
\vspace{-2mm}
\begin{equation}
    \label{eq:c}
    C = \underset{C}{\arg\min} \sum_{i=1}^{k} \sum_{n \in c_i} d(n, \mu_i)    
\end{equation}
where \(d(n, \mu_i)\) is a distance function that measures the dissimilarity between a node \(n\) and the centroid \(\mu_i\) of cluster \(c_i\), reflecting the fit between node capabilities and workflow requirements.

\subsubsection{$k$-means clustering as a VEC scheduling solution}
To address the challenge of effectively clustering VEC nodes, we have developed an advanced mechanism utilizing the $k$-means algorithm~\cite{likas2003global}. As illustrated in Fig.~\ref{fig:vec-solution}, our VECA solution architecture incorporates a clustering feature to cluster VEC node resources based on capacity characteristics. Once clusters are defined, we initiate the first stage of our two-phase scheduling mechanism, selecting the cluster that is more likely effective and suitable to execute a particular workflow. We limit the search granularity of VEC nodes at the cluster level, reducing search latency times associated with this phase. 

Details of this approach are presented later in Section~\ref{sec:methodology}.

\subsection{Managing Dynamic Volunteer Resources}

\subsubsection{Challenges}
VEC environments are characterized by the volatility of volunteer-provided resources, which manifest in unpredictable availability patterns. In Fig.~\ref{fig:vec-solution}, we illustrate this characteristic where in \textit{FaaS Cluster n} there occurs an event where a VEC node instantly goes offline in the middle of workflow execution. This intermittency poses a substantial risk to the continuity and reliable/predictable execution of scientific workflows~\cite{edgeenabler, VECreqchallim}. Formally, the challenge is to develop a predictive and adaptive system that minimizes the disruptive impact of this intermittency on workflow execution within the VEC environment. This requires both forecasting future availability, and a fail-over mechanism to deal with smooth recovery and continued operations following failures.

\paragraph{Modeling Node Availability}
Node availability can be modeled as a stochastic process, where the state of each node is represented as a binary variable \( x_t \) at time \( t \), indicating whether the node is online (\( x_t = 1 \)) or offline (\( x_t = 0 \)).

\paragraph{Predictive Modeling}
We propose using a Recurrent Neural Network (RNN) to model the time-dependent sequences of node availability. The state of the RNN at time \( t \), denoted as \( h_t \), is computed as:
\vspace{-2mm}
\begin{equation}
    \label{eq:relu}
    h_t = \text{ReLU}(W_{ih} x_t + b_{ih} + W_{hh} h_{(t-1)} + b_{hh})
\end{equation}
where \( W_{ih} \), \( W_{hh} \) are the input-hidden and hidden-hidden weight matrices, \( b_{ih} \), \( b_{hh} \) are biases, and ReLU is the activation function providing non-linearity.

The objective is to maximize the overall availability of the computing resources by minimizing the probability of workflow failures due to VEC node unavailability. This is achieved by optimizing the selection and scheduling strategies based on the predictive models and fail-over mechanisms as detailed later in Section~\ref{sec:2-phasescheduler}.

\subsubsection{Solution to address the dynamic nature of volunteer resources}
\label{sec:proposed_solution}
We propose a nuanced assessment of nodes within the selected cluster, focusing on their future availability, and geographic proximity. We propose a two-phase scheduler as shown in Fig.~\ref{fig:vec-solution} by harnessing a RNN-based time-series forecasting and fail-over mechanism. In the event of an execution interruption, the system is adept at dynamically reassigning the workflow to the subsequent optimal node within the cluster by reading the workflow details from the cluster cache. This prevents having to go to the source for assignment, reducing the round trip times and maintaining a seamless operational flow. Such a process design with a Two-phase scheduling approach aims to ensure an efficient, reliable, and interruption-resistant scientific workflow execution. Details of this approach are presented later in Section~\ref{sec:2-phasescheduler}.

\vspace{-1mm}
\section{Capacity Based $k$-means Clustering}
\label{sec:methodology}

In this section, we detail the steps and related implementation of our $k$-means clustering approach for intricate VEC resource management and scientific workflow orchestration. 

\vspace{-2mm}
\subsection{VEC Nodes Characterization, Optimization and Clustering}
\label{Node Characteristics}

We consider the capacity characteristics of VEC nodes, recognizing them as crucial components for scientific workflow execution. This characterization encompasses quantitative metrics such as: (a) number of CPUs, representing the processing power, (b) RAM to indicate the memory capacity of each node, and (c) storage size to reflect the available storage on each node.
We utilized the Elbow method to determine the optimal number of clusters from the given pool of VEC resources. In our example case that involved 50 VEC nodes, the Elbow method results in 4 clusters. 

\vspace{-2mm}
\subsection{Implementing $k$-means for VEC Node Clustering}

For our $k$-means clustering implementation, we trained our model on 50 VEC nodes, generating their characteristics, mentioned in Section \ref{Node Characteristics}. This dataset was generated to replicate the real-world scenario of a VEC computing environment. Before starting the clustering process, we standardized the dataset using the \textit{StandardScaler} from \textit{scikit-learn}, ensuring that each feature had mean of approximately $0$, and variance of $1$. Standardization is a crucial pre-processing step, especially when features have different units of measurement, as it puts them on the same scale allowing the clustering algorithm to converge more effectively. we have used heuristic Elbow method to determine the optimal number of clusters. This approach involves running the $k$-means clustering process on the dataset for a range of values of $k$ (the number of clusters). In this case, we consider $k= range(1, 9)$ for the experimentation. For each value of $k$, the Sum of Squared Distances (SSD) within each cluster is computed. This measure, also known as inertia, quantifies the compactness of the clusters, with lower values indicating better clustering.

\begin{algorithm}[h]
    \scriptsize
    \caption{\footnotesize {Determining Optimal Number of Clusters using Elbow Menthod and Clustering VEC Nodes using $k$-means}}
    \label{alg:K_Means}
    \SetAlgoLined
    $Sum\_of\_squared\_distances \gets [];$\\
    $K \gets \text{range}(1, 9);$\\
    $X \gets df\_encoded.\text{values}$ \tcp*{Load the dataset}
    $X\_s \gets \text{StandardScaler().fit\_transform}(X)$ \tcp*{Standardize data} 
    \For{$num\_clusters \text{ in } K$}{
        $kmeans \gets \text{KMeans}(n\_clusters=num\_clusters);$ \\
        $kmeans.\text{fit}(X\_s);$ \\
        $Sum\_of\_squared\_distances.\text{append}(kmeans.\text{inertia\_});$ \\
    }
    $\texttt{Plot\_Elbow\_Curve}(K, Sum\_of\_squared\_distances);$
    
    $kmeans \gets \text{KMeans}(n\_clusters=4)$ \tcc*{Initialize KMeans with optimal number of clusters from Elbow method}
    $kmeans.\text{fit}(X\_s);$ \\
    $labels \gets kmeans.\text{predict}(X\_s);$
\end{algorithm}
\begin{figure}[t] 
    \centering
    \vspace{-3mm}
    \includegraphics[width=0.78\linewidth]{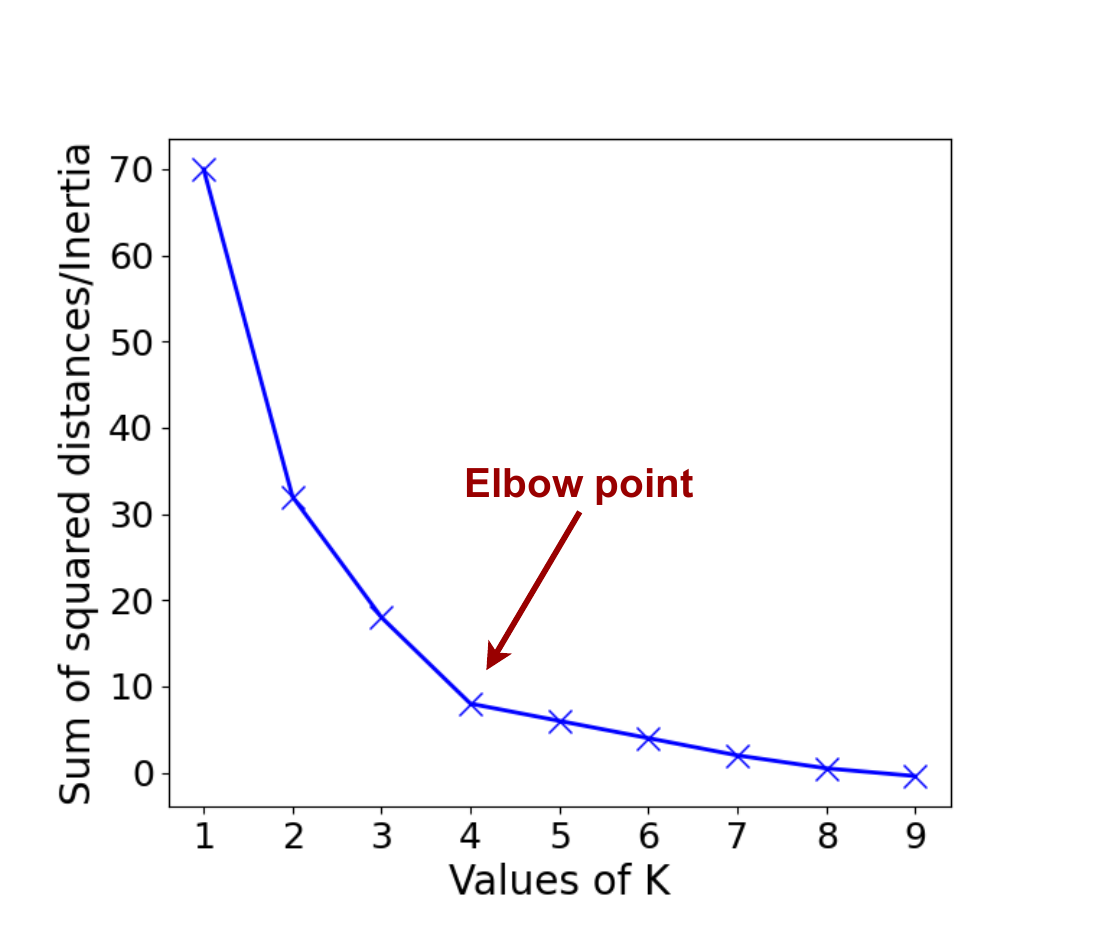}
    \caption{Elbow Plot to determine optimal number of $k$ clusters.}
    \label{fig:ElbowPlot}
    \vspace{-5mm}
\end{figure}

The $k$-means volunteer node clustering algorithm (as depicted by Algorithm~\ref{alg:K_Means}) was applied to the dataset containing multidimensional descriptions of the VEC nodes. The SSD for each value of $k$ was plotted against the corresponding $k$ values to visualize the Elbow curve as shown in Fig.~\ref{fig:ElbowPlot}. The ``Elbow'' point on this curve, where the rate of decrease sharply changes, indicates the appropriate number of clusters for the data. This is based on the principle that increasing the number of clusters beyond the true number does not significantly improve the SSD. This method is particularly useful for identifying the value of $k$ that balances informativeness with simplicity, thereby avoiding over-fitting. The Elbow plot depicted in Fig.~\ref{fig:ElbowPlot} helps in determining the number of clusters where the additional variance explained does not justify adding another cluster. In this specific case, the optimal number of clusters is $4$. With the VEC nodes appropriately grouped based on their similarity in capacity characteristics, re-clustering is performed when ever there is a 10\% increase in the number of cluster nodes. Following this, the VEC environment is now prepared for the second phase of our approach.

\begin{algorithm}[h]
    \scriptsize
    \caption{\footnotesize {Two-Phase Scheduler for VEC Resource Allocation}}
    \label{alg:2-phase-scheduler}
    \SetAlgoLined
    
    \KwIn{Workflow $W$ from VEC user, Clusters $C$, Nodes $N$ in each cluster}
    \KwOut{Execution Status and Result Delivery}
    
    \SetKwFunction{FSelectCluster}{SelectCluster}
    \SetKwFunction{FPredictNodeAvailability}{PredictNodeAvailability}
    \SetKwFunction{FSelectNearestNode}{SelectNearestNode}
    \SetKwFunction{FExecuteWorkflow}{ExecuteWorkflow}
    \SetKwFunction{FReturnResults}{ReturnResults}
    \SetKwFunction{FVECWorkflowScheduler}{VECWorkflowScheduler}
    
    \SetKwProg{Fn}{Function}{:}{}
    
    \Fn{\FSelectCluster{$W, C$}}{
        $selectedCluster \gets$ cluster with capacity closest to $W.capacity$\;
        Enqueue $W$ in $selectedCluster.queue$\;
        \Return $selectedCluster$\;
    }
    \Fn{\FPredictNodeAvailability{$cluster, W$}}{
        $nodeList \gets$ list all available nodes in $cluster$. If $W$ needs to be executed in confidential computing mode, filter VEC nodes that support confidential computing\;
        $availabilityList \gets$ []\;
        \For{each node $n$ in $nodeList$}{
            $availability \gets RNN\_Predict(n, W)$\;
            Append $(n, availability)$ to $availabilityList$\;
        }
        $orderedNodes \gets$ sort $availabilityList$ by predicted\_availability descending\;
        Store $W$ and $orderedNodes$ in Redis cache\;
        \Return $orderedNodes$\;
    }
    \Fn{\FSelectNearestNode{$OrderedNodes$}}{
        $eligibleNodes \gets$ filter nodes from $OrderedNodes$ with predicted\_availability $\geq 0.8$\;
        \If{empty $eligibleNodes$}{
            $selectedNode \gets$ top node from $OrderedNodes$\;
        }
        \Else{
            $selectedNode \gets$ nearest node from $eligibleNodes$ to VEC user\;
        }
        \Return $selectedNode$\;
    }
    \Fn{\FExecuteWorkflow{$Node, W$}}{
        Execute $W$ on $Node$ using FaaS\;
        \If{execution fails}{
            Retrieve $W$, $orderedNodes$ from Redis cache\;
            $Node \gets \FSelectNearestNode{orderedNodes}$\;
            Go to \FExecuteWorkflow{$Node, W$}\;
        }
        \Return Execution success\;
    }
    \Fn{\FReturnResults{$Node, W$}}{
        $results \gets$ collect results from $Node$\;
        Send $results$ to Main scheduler in phase one\;
        Store the results in Flask server\;
        Display $results$ on User UI\;
    }
    \Fn{\FVECWorkflowScheduler{$W$}}{
        $cluster \gets \FSelectCluster{W, C}$\;
        $orderedNodes \gets \FPredictNodeAvailability{cluster, W}$\;
        $executionNode \gets \FSelectNearestNode{orderedNodes}$\;
        $success \gets \FExecuteWorkflow{executionNode, W}$\;
        \If{$success$}{
            \FReturnResults{executionNode, W}\;
        }
    }
\end{algorithm}
\vspace{-4mm}

\begin{figure}[h]
\centering
\vspace{-5mm}
\includegraphics[width = 0.92\linewidth]{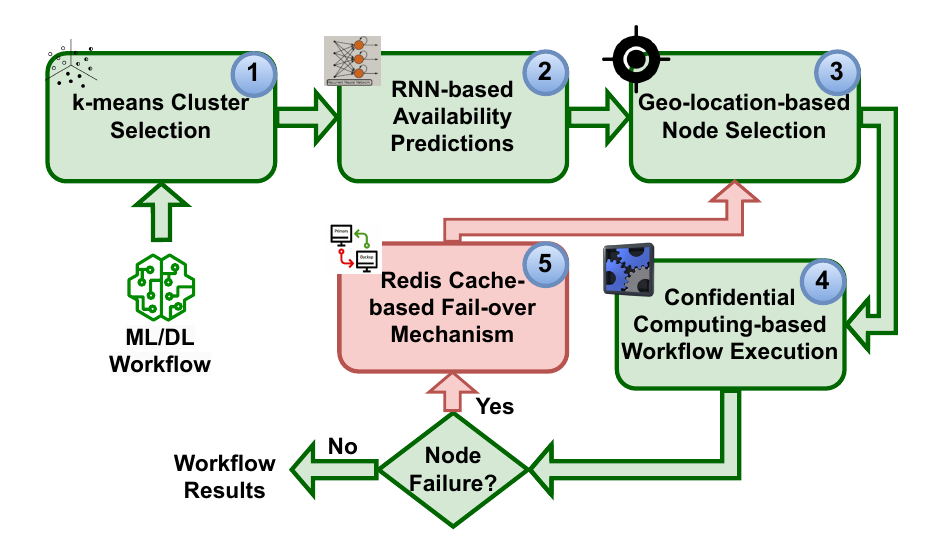}
\caption{\footnotesize{Pipeline for the two-phase scheduler.  }}
\label{fig:sch-pipeline}
\vspace{-5mm}
\end{figure}

\vspace{-1mm}
\section{Distributed Two-phase scheduler}
\label{sec:2-phasescheduler}

Scientific users will be provided with an User Interface to submit the workflow. Upon workflow submission, the scheduler initiates a two-phase scheduling algorithm (as shown in Algorithm~\ref{alg:2-phase-scheduler}) as a pipeline~\ref{fig:sch-pipeline}. The phase one of the scheduler is executed in the Cluster Selection Controller node of the Cloud Hub. In the phase one of the pipeline, the scheduler selects a cluster based on the workflow's capacity requirements using $k$-means algorithm by passing the new data point as an input to the model to determine the cluster that is nearest to the new data point (as depicted in Step 1 in Fig.~\ref{fig:sch-pipeline}). This involves delegating the workflow to a cluster agent node, which possesses comprehensive data on VEC nodes availability in that particular cluster. Phase two of the scheduler is executed in the Agent Node of the selected cluster and the asynchronous inter-scheduler communication takes place using RabbitMQ message queue. In phase two, the selected cluster's nodes undergo evaluation based on future availability, and geo-location. This process utilizes an RNN-based feed-forward neural network trained on time-series data to forecast node availability (Step 2). The scheduler then assigns the workflow to the most suitable node, taking into account geographic proximity to the scientific user for node selection (Step 3). Below, we provide a detailed implementation involved in the two-phase scheduler.

\vspace{-1mm}
\subsection{RNN-based Time-series Forecasting for Availability Prediction}
In VECA, time series forecasting serves as a critical tool for enhancing system robustness by predicting the availability of VEC nodes. It allows for preemptive scheduling decisions, ensuring that the workflows are allocated to nodes when they are most likely to be available, thereby minimizing down times and optimizing the resource utilization. This is Step 2 in the Fig.~\ref{fig:sch-pipeline}, where we first sample all the available VEC nodes of the cluster at a given moment and pass them through an RNN model to predict the future availability. RNN-based feed-forward neural networks, with their inherent strength in handling sequential data, are an ideal choice for this forecasting task~\cite{rnn-feedforward}. Their ability to learn from historical availability patterns enables the prediction of future node statuses, making the system more reliable and efficient. To the best of our knowledge, this is the first paper to propose time series forecasting for VEC computing. Herein, we further detail the model implementation.

\subsubsection{Custom dataset preparation}
To evaluate the effectiveness of our approach in a realistic setting, we constructed a synthetic dataset encompassing data for 50 VEC nodes and their availability over a one-year period. This dataset incorporates diverse availability patterns, reflecting real-world scenarios.  Some nodes exhibit limited availability during typical working hours (weekdays, 9AM-5PM), while others, likely contributed by research labs or universities, demonstrate high availability throughout the week. The dataset enables the model to learn the relationships between day of the week, hour, and VEC node ID, ultimately predicting availability with robustness. This approach can be readily extended to capture real-world VEC node availability data using node monitoring.

\subsubsection{Data pre-processing}
As part of the pre-processing, categorical features (VolunteerID, Weekday) are converted into a numerical format using OneHotEncoder. This step expands the dimensionality, where each unique category is represented by a binary vector. The `Hour' feature undergoes normalization via StandardScaler, transforming it to have a mean of $0$ and a standard deviation of $1$, improving model convergence speed and stability.

\subsubsection{Model architecture}
The RNN model is constructed with a specified input size (matching the feature vector's dimension), hidden size (determining the complexity and capacity of the model), and output size ($1$, for binary availability prediction). RNNs leverage the sequential nature of time series data, using the hidden state that carries information across time steps to capture temporal dependencies.

\noindent The input encoding format for RNN is given by:
\vspace{-2mm}
\begin{equation}
\label{eq:encoding}
\begin{aligned}
X = [\texttt{OneHot}(\textit{VID}, \textit{WD}), \texttt{StandardScaler}(\textit{H})]
\end{aligned}
\end{equation}
where \textit{VID, WD} and \textit{H} are VolunteerID, Weekday, and Hour respectively.

\noindent The hidden state at time \(t\) is computed as:
\vspace{-2mm}
\begin{equation}
    \label{eq:hidden}
    h_t = \tanh(W_{ih} x_t + b_{ih} + W_{hh} h_{(t-1)} + b_{hh})
\end{equation}

\noindent The output at time \(t\) is given by:
\vspace{-2mm}
\begin{equation}
    \label{eq:output}
    o_t = W_{ho} h_t + b_o
\end{equation}
\noindent The predicted availability is obtained using the sigmoid function:
\vspace{-2mm}
\begin{equation}
    \label{eq:sigmoid}
    \hat{y}_t = \sigma(o_t)
\end{equation}
where \(\sigma\) denotes the sigmoid function, transforming the RNN output to a probability for availability prediction.

In the provided Equations~\ref{eq:hidden},~\ref{eq:output}, and~\ref{eq:sigmoid}, the values \(W_{ih}\), \(W_{hh}\), and \(W_{ho}\) represent the weight matrices for transitions from input to hidden layer, hidden layer to itself, and hidden layer to output layer, respectively. The bias terms for these transitions are denoted by \(b_{ih}\), \(b_{hh}\), and \(b_{o}\) respectively. The \(\tanh\) function in Equation~\ref{eq:hidden} introduces non-linearity to the hidden state computation, while the sigmoid function \(\sigma\) transforms the RNN's output to a probability, suitable for binary classification tasks such as availability prediction where the value ranges from $0$ to $1$, depicting the probability of VEC node availability for a specific time under consideration.

\subsubsection{Training process}
We have trained the dataset using $60$ epochs and $128$ hidden layers, where the model makes predictions, calculates loss via a \texttt{BCEWithLogitsLoss} loss function, combining logistic regression with binary cross-entropy loss, and updates weights using back propagation with an Adam optimizer for adaptive learning rate adjustments finalized to $0.001$. The RNN’s forward pass computes the output considering current input and the previous time step's hidden state, followed by linear transformation for final prediction.

\subsubsection{Output interpretation}
The output generated by the trained model indicates the probability of a node remaining online, with values scaled between $0$ and $1$. A value approaching $1$ suggests a high likelihood of the node maintaining availability for time $t$. This probabilistic output enables a nuanced assessment of node reliability in real-time scenarios.

\vspace{-1mm}
\subsection{Geo-location-based Node Selection for Workflow Execution}
Incorporating geo-location awareness into the system significantly enhances user satisfaction by prioritizing the selection of computing nodes closest to the user's location for workflow execution. This is Step 3 in Fig.~\ref{fig:sch-pipeline} where we filter the \textit{predicted\_availability} of VEC nodes \(\geq\) $0.8$ and pick the nearest VEC node for executing the workflow. By leveraging geographical proximity, the system can offer more responsive and tailored computing services. This geo-location-based selection strategy, underpinned by mathematical distance calculation as illustrated in Algorithm~\ref{alg:geo-location}, is pivotal for optimizing resource allocation in distributed computing.

\subsection{Confidential Computing-based Workflow Execution}
\label{sec:aws-nitro-cc-approach}

As the next step (Step 4) in the Two-phase scheduler, if the scientific user chooses to run the workflow on a TEE that delivers CC, the workflow will be assigned to the VEC node that has AWS Nitro installed. In the implementation of CC using AWS Nitro, the process is structured into four distinct steps, ensuring the integrity and confidentiality of the computations. 

\paragraph{Building enclave} Involves building the Encrypted Image Snapshot (EIS) from the Docker Image safeguarding it during storage and transit.

\paragraph{Running enclave} Involves running the enclave on AWS Nitro enabled EC2 instances, this provides isolated CPU and memory resources that are accessible only to the enclave itself

\paragraph{Validating enclave} This is achieved through the Attestation Document, a cryptographic proof generated at the enclave's startup, detailing its identity and confirming the integrity of its contents. 

\paragraph{Terminating enclave} Once the required computations are completed, the enclave is securely shut down, ensuring that all sensitive data and state information are erased, preventing any residual data exposure.

Through these steps, we adapt the AWS Nitro services for executing workflows in a secure and controlled manner, utilizing advanced isolation, encryption, and attestation to meet the stringent demands of confidential computing.

\subsection{Fail-over Mechanism}
\label{sec:fail-over}

In the event of a workflow execution failure on any VEC node, the system's fail-over mechanism plays a crucial role in ensuring robust and efficient recovery. This process leverages the Redis cache to swiftly retrieve essential workflow details and the pre-computed order of VEC nodes, thereby avoiding the need to revisit the origin of the workflow data or to re-execute the RNN model for node prioritization. By storing this data in the Redis cache, the system significantly reduces round trip times and avoids the computational overhead associated with re-running the initial phases of the scheduler. As depicted in Step 5 of Fig.~\ref{fig:sch-pipeline}, upon failure, the process resumes from Step 3, seamlessly continuing the execution without unnecessary delays. The resultant workflow data is then promptly relayed back to the agent node, subsequently forwarded to the main scheduler, and finally stored on a Flask server to display the execution results to the scientific workflow user. This fail-over governance strategy not only enhances the system's resilience against disruptions but also ensures that the resource allocation remains optimal i.e., execution times are minimized, maintaining a high level of service continuity for end users.

\section{Performance Evaluation}
\label{sec:experimentation_results}

We have developed a comprehensive VEC web-based tool published on GitHub~\cite{VECA}, where a scientific workflow user can submit his/her workflow using a provided user interface. To implement the VECA solution for evaluation experiments, we define a technology stack that includes OpenFaaS, MicroK8s, and Dockerization. OpenFaaS enables encapsulation of complex functionalities into scalable, serverless functions, which are ideally suited for the heterogeneous VEC environments. MicroK8s simplifies Kubernetes orchestration, offering a lightweight solution ideal for the decentralized nature of the VEC resources. Through these technologies, we ensure that our system not only addresses current security challenges but also is amenable to adapt to the evolving landscape of distributed computing.

In the following, we detail our evaluation experiments on our approach using VEC Node Search Latency, Productivity Rate metrics.

\vspace{-2mm}
\subsection{VEC Node Search Latency}
VEC Node Search Latency is a crucial performance metric in VEC environments, as it measures the time taken to identify the most appropriate VEC node for executing a given workflow. Lower latency is indicative of a more efficient system, contributing to faster workflow deployment and execution, which is critical in time-sensitive scientific computations. Thus, we study the performance of our approach for VEC node search latency and compare with state-of-the-art methods i.e., VELA~\cite{vela} and VECFlex~\cite{vecflex}.

In VECFlex, the entire pool of nodes, which can be substantial in number, must be sampled to identify the optimal node for task execution. This process is defined by:
\vspace{-2mm}
\[
\text{Latency}_{\text{VECFlex}} = \text{Time}_{\text{Node Sampling}}(n),
\]
where $n$ is the total number of nodes. This exhaustive search, while thorough, introduces significant latency, making it less desirable for time-critical tasks.

VELA, on the other hand, categorizes nodes into clusters. When a workflow is submitted, VELA randomly selects a subset of clusters and then samples nodes from these clusters. This introduces randomness and potential inefficiencies into the node selection process:
\vspace{-2mm}
\[
\text{Latency}_{\text{VELA}} = \text{Time}_{\text{Cluster Selection}} + \text{Time}_{\text{Node Sampling}}(n \cdot c).
\]

where $n$ is the number of nodes per cluster and $c$ is the number of clusters sampled. Although the search space is reduced, when compared to VECFlex, the random selection of clusters does not guarantee that the chosen VEC nodes are best suited for the workflow requirements as VEC node characteristics are not considered.

Our approach, VECA, optimizes the process of VEC node search by intelligently selecting a cluster that closely matches the workflow's capacity requirements. Consequently, only the VEC nodes within this single cluster are sampled:
\vspace{-2mm}
\[
\text{Latency}_{\text{VECA}} = \text{Time}_{\text{Cluster Selection}} + \text{Time}_{\text{Node Sampling}}(n).
\]

Although there is an additional computational overhead for selecting the most suitable cluster, VECA's targeted approach significantly reduces the overall search space by narrowing the search to VEC nodes within a single, capacity-matched cluster. In addition, VECA reduces the VEC node search latency while maintaining a high probability of node suitability for the task requirements. This fine-grained and predictive scheduling approach exemplifies the optimization of resource allocation within VEC systems, thus balancing efficiency and precision in task scheduling.
\begin{figure}[t]
\centering
\vspace{-4mm}
\includegraphics[width = 0.7\linewidth]{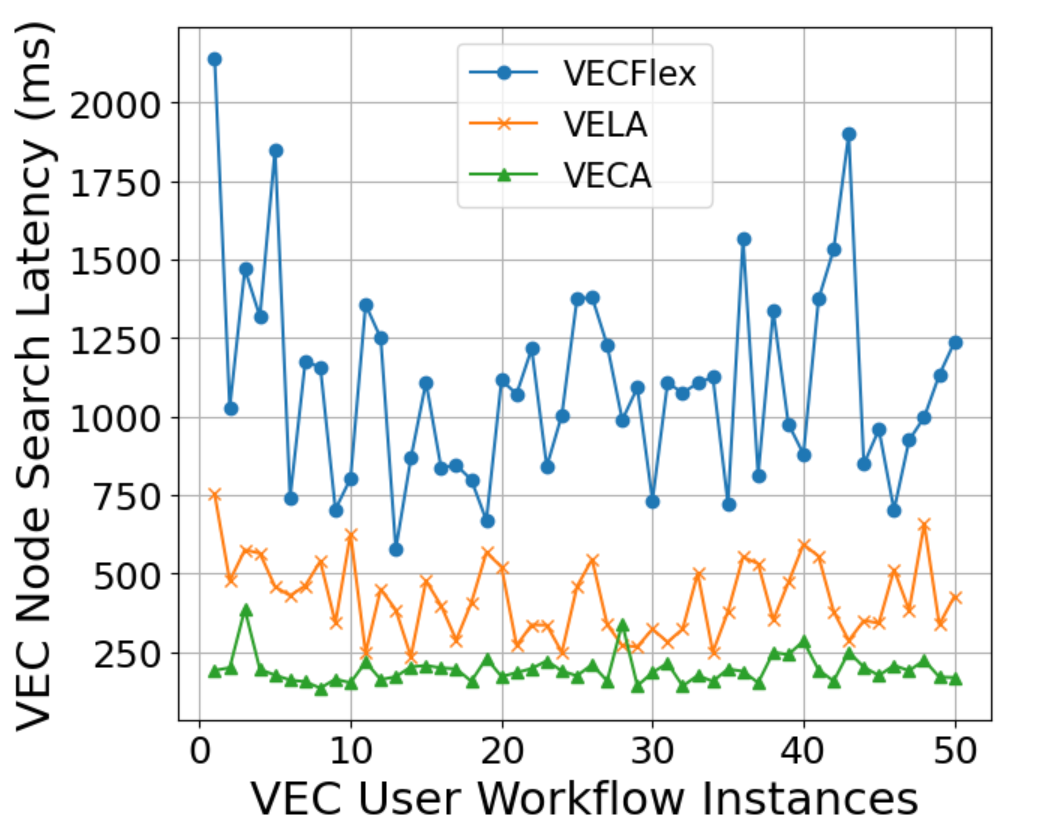}
\caption{\footnotesize{Results on VEC node search latency across 
 50 workflow instances.
}
}
\label{fig:latency_plot}
\vspace{-6mm}
\end{figure}

To validate the efficiency of our VECA system against state-of-the-art methods such as VELA and VECFlex, we implemented a simulation within a structured VEC environment consisting of 50 VEC nodes, strategically divided into 4 clusters using the $k$-means algorithm. We conducted experiments by scheduling 50 workflow instances under varied workload conditions. As illustrated in Fig.~\ref{fig:latency_plot}, the results demonstrate a consistently low node search latency for VECA compared to VELA and VECFlex. The graph reveals that, generally, VECA achieves lower latency in task execution, which underscores the system's effectiveness in optimizing VEC node search within clusters. Notably, there are instances where the latency numbers for VELA approach those of VECA. This convergence typically occurs during periods when multiple VEC nodes are engaged in other tasks, limiting the pool of immediately available VEC nodes. In such scenarios, VECA and VELA are restricted to selecting from a similar subset of freely available VEC nodes, which momentarily equalizes their performance.

\begin{figure}[h]
\centering
\vspace{-4mm}
\includegraphics[width = 0.82\linewidth]{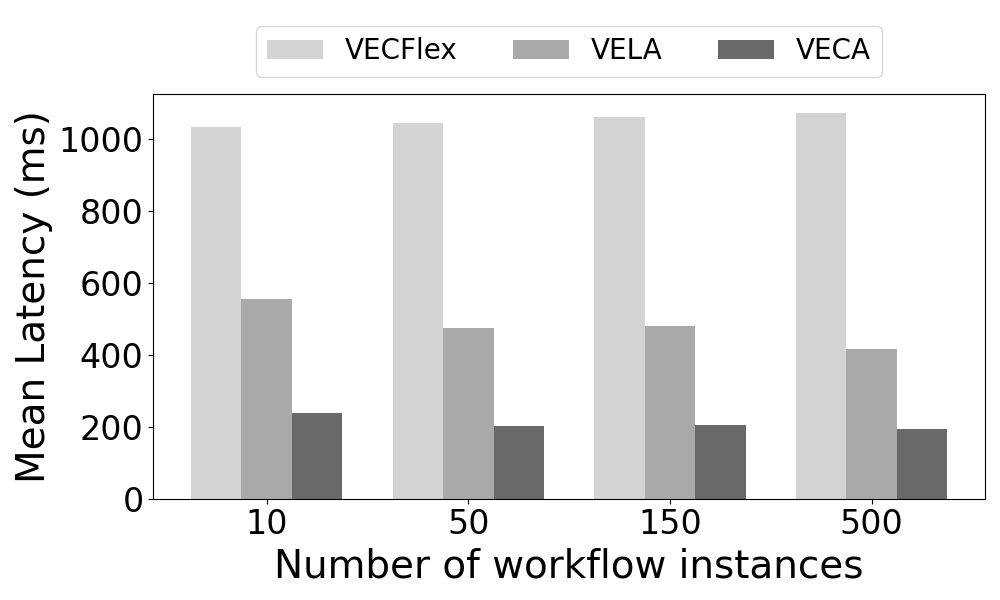}
\caption{\footnotesize{Performance of the different approaches over a varying number of workflow instances.
}
}
\label{fig:latency_histo}
\vspace{-3mm}
\end{figure}

VECA consistently outperforms the state-of-the-art solutions over a broad range of scales. Specifically, we performed experiments for variable set of workflow instances with increasing scale \{10, 50, 150, 500\}, as shown in Fig.~\ref{fig:latency_histo}, highlighting its superior efficiency in VEC Node Search under distributed workloads. We can note that our VECA consistently exhibits a two-fold reduction in VEC node search latency compared to the next best solution i.e., VELA. The observed performance advantage is primarily due to VECA's intelligent clustering and node selection algorithms, which significantly reduce unnecessary computational overheads for sampling VEC nodes in the resource allocation processes, ensuring optimal resource allocation and faster response times in dynamic VEC environments.

\vspace{-2mm}
\subsection{Productivity Rate}
The productivity rate metric is used to measure the efficiency of a system in successfully recovering from failures and continuing operation without significant loss of functionality or data. In the context of VEC computing environments, it could refer to the system's capability to handle VEC node failures by quickly resuming tasks on alternative VEC nodes, thus ensuring minimal disruption and maintaining system performance. This metric is particularly important in distributed systems where tasks are critical and require high availability.

We define the productivity rate as the proportion of the total execution time that was not taken up by recovery actions, expressed as a percentage. This measure indicates the efficiency of the recovery process—a higher productivity rate indicates a more resilient system.
\vspace{-2mm}
\[
\text{Productivity Rate} = \left(1 - \frac{\textit{Time Taken for Recovery}}{\textit{Total Execution Time}}\right) \times 100\%,
\]
where:
\begin{itemize}
  \item \textit{Time Taken for Recovery} is the duration from the onset of a failure to the resumption of normal operations.
  \item \textit{Total Execution Time} is the sum of the recovery time and any time spent on normal operations as part of the workflow execution.
\end{itemize}

\begin{figure}[t]
\centering
\vspace{-4mm}
\includegraphics[width = 0.85\linewidth]{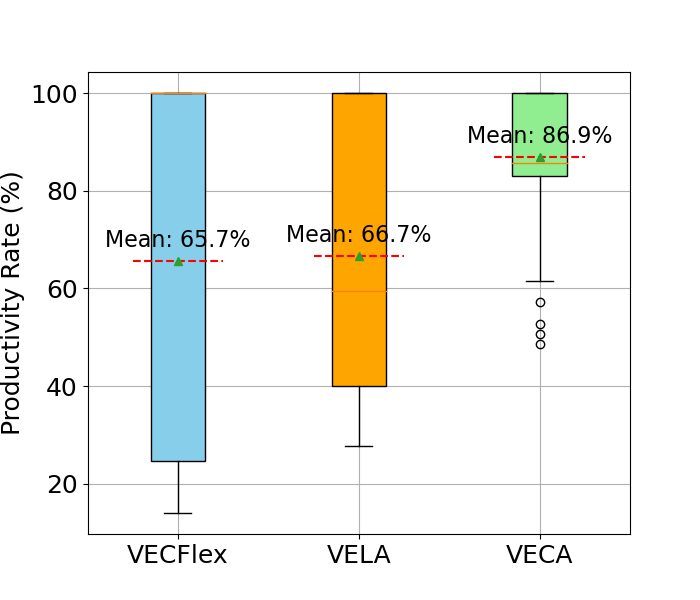}
\vspace{-5mm}
\caption{\footnotesize{Results on Productivity Rate across 50 workflow instances.}
}
\label{fig:recoveryrate_boxplot}
\vspace{-6mm}
\end{figure}
Our experimentation results, illustrated through a box plot analysis as shown in Fig.~\ref{fig:recoveryrate_boxplot}, demonstrate that VECA significantly outperforms both state-of-the-art solutions i.e., VECFlex and VELA in terms of productivity rates. The mean productivity rate for VECA was 86.9\%, compared to 66.7\% for VELA and 65.7\% for VECFlex. This superior performance of VECA can be attributed to its advanced availability prediction mechanism coupled with a strategic caching system empowered by Redis Cache, which collectively ensures a rapid resumption of workflow tasks post-failure without the need for re-sampling of nodes. By adopting VECA, VEC environments can achieve higher resilience and reliability, thus broadening their applicability in critical ML/DL-based scientific workflows e.g.., bioinfomatics and health informatics, where downtime of VEC nodes can have significant impacts on the expected productivity in terms of execution times.

\vspace{-2mm}
\section{Conclusion and Future Works}
\label{sec:conclusion_futureworks}
In this paper, we proposed a solution viz., VECA for reliable and confidential resource clustering for VEC computing in order to address the challenges of managing VEC resources for ML/DL-based scientific workflows. By implementing capacity-based clustering, confidential computing integration, and globally distributed scheduling schemes, VECA significantly improves the ability to recover from VEC node failures, and offers a systematic set of protections to ensure privacy preservation of the ML/DL-based scientific workflows in VEC computing environments. The evaluation results demonstrate the effectiveness of VECA in reducing VEC node search latency in identifying optimal VEC nodes for workflow execution, and enhancing productivity rates to complete workflow executions, compared to existing state-of-the-art solutions such as VECFlex and VELA.

Future research can focus on integrating federated machine learning to create cluster capacities suitable for other diverse scientific workflows e.g., medical imaging with unique performance and privacy preservation requirements.

\bibliographystyle{IEEEtran} 
\bibliography{Mybib.bib}

\begin{thebibliography}{10}
\providecommand{\url}[1]{#1}
\csname url@samestyle\endcsname
\providecommand{\newblock}{\relax}
\providecommand{\bibinfo}[2]{#2}
\providecommand{\BIBentrySTDinterwordspacing}{\spaceskip=0pt\relax}
\providecommand{\BIBentryALTinterwordstretchfactor}{4}
\providecommand{\BIBentryALTinterwordspacing}{\spaceskip=\fontdimen2\font plus
\BIBentryALTinterwordstretchfactor\fontdimen3\font minus
  \fontdimen4\font\relax}
\providecommand{\BIBforeignlanguage}[2]{{%
\expandafter\ifx\csname l@#1\endcsname\relax
\typeout{** WARNING: IEEEtran.bst: No hyphenation pattern has been}%
\typeout{** loaded for the language `#1'. Using the pattern for}%
\typeout{** the default language instead.}%
\else
\language=\csname l@#1\endcsname
\fi
#2}}
\providecommand{\BIBdecl}{\relax}
\BIBdecl

\bibitem{seti}
E.~Korpela, D.~Werthimer, D.~Anderson, J.~Cobb, and M.~Leboisky,
  ``Seti@home-massively distributed computing for seti,'' \emph{Computing in
  Science \& Engineering}, vol.~3, no.~1, pp. 78--83, 2001.

\bibitem{vecflex}
M.~L. Alarcon, M.~Nguyen, A.~Pandey, S.~Debroyand, and P.~Calyam, ``Vecflex:
  Reconfigurability and scalability for trustworthy volunteer edge-cloud
  supporting data-intensive scientific computing,'' in \emph{2022 IEEE/ACM 15th
  International Conference on Utility and Cloud Computing (UCC)}, 2022, pp.
  151--156.

\bibitem{edgeenabler}
T.~Mengistu, A.~Albuali, A.~Alahmadi, and D.~Che, \emph{Volunteer Cloud as an
  Edge Computing Enabler}, 06 2019, pp. 76--84.

\bibitem{BOINC}
``Boinc, compute for science,'' \url{https://boinc.berkeley.edu/}, Accessed
  April 2024.

\bibitem{kubeedge}
``Kubeedge,'' \url{https://kubeedge.io/}, Accessed April 2024.

\bibitem{osdf}
``Open science data federation,'' \url{https://osg-htc.org/services/osdf},
  Accessed April 2024.

\bibitem{fog-edge-resmngmnt-survey}
\BIBentryALTinterwordspacing
C.-H. Hong and B.~Varghese, ``Resource management in fog/edge computing: A
  survey on architectures, infrastructure, and algorithms,'' \emph{ACM Comput.
  Surv.}, vol.~52, no.~5, sep 2019. [Online]. Available:
  \url{https://doi.org/10.1145/3326066}
\BIBentrySTDinterwordspacing

\bibitem{survey-iot-edge}
\BIBentryALTinterwordspacing
P.~Gkonis, A.~Giannopoulos, P.~Trakadas, X.~Masip-Bruin, and F.~D’Andria, ``A
  survey on iot-edge-cloud continuum systems: Status, challenges, use cases,
  and open issues,'' \emph{Future Internet}, vol.~15, no.~12, 2023. [Online].
  Available: \url{https://www.mdpi.com/1999-5903/15/12/383}
\BIBentrySTDinterwordspacing

\bibitem{vela}
\BIBentryALTinterwordspacing
T.~Pusztai, S.~Nastic, P.~Raith, S.~Dustdar, D.~Vij, and Y.~Xiong, ``Vela: A
  3-phase distributed scheduler for the edge-cloud continuum,'' in \emph{2023
  IEEE International Conference on Cloud Engineering (IC2E)}.\hskip 1em plus
  0.5em minus 0.4em\relax Los Alamitos, CA, USA: IEEE Computer Society, sep
  2023, pp. 161--172. [Online]. Available:
  \url{https://doi.ieeecomputersociety.org/10.1109/IC2E59103.2023.00026}
\BIBentrySTDinterwordspacing

\bibitem{aws-nitro}
``Aws nitro enclaves,'' \url{https://aws.amazon.com/ec2/nitro/nitro-enclaves/},
  Accessed April 2024.

\bibitem{OpenFaaS}
``Openfaas, serverless functions, made simple,''
  \url{https://www.openfaas.com/}, Accessed April 2024.

\bibitem{MicroK8s}
``Microk8s,'' \url{https://microk8s.io/}, Accessed April 2024.

\bibitem{g2pdeep}
\BIBentryALTinterwordspacing
S.~Zeng, Z.~Mao, Y.~Ren, D.~Wang, D.~Xu, and T.~Joshi, ``{G2PDeep: a web-based
  deep-learning framework for quantitative phenotype prediction and discovery
  of genomic markers},'' \emph{Nucleic Acids Research}, vol.~49, no.~W1, pp.
  W228--W236, 05 2021. [Online]. Available:
  \url{https://doi.org/10.1093/nar/gkab407}
\BIBentrySTDinterwordspacing

\bibitem{pasml}
\BIBentryALTinterwordspacing
H.~Salah and S.~Srinivas, ``Predict, then schedule: Prescriptive analytics
  approach for machine learning-enabled sequential clinical scheduling,''
  \emph{Computers \& Industrial Engineering}, vol. 169, p. 108270, 2022.
  [Online]. Available:
  \url{https://www.sciencedirect.com/science/article/pii/S0360835222003357}
\BIBentrySTDinterwordspacing

\bibitem{Redis}
``Redis,'' \url{https://redis.io/}, Accessed April 2024.

\bibitem{clara}
\BIBentryALTinterwordspacing
S.~Gonzalo, J.~M. Marquès, A.~García-Villoria, J.~Panadero, and L.~Calvet,
  ``Clara: A novel clustering-based resource-allocation mechanism for
  exploiting low-availability complementarities of voluntarily contributed
  nodes,'' \emph{Future Generation Computer Systems}, vol. 128, pp. 248--264,
  2022. [Online]. Available:
  \url{https://www.sciencedirect.com/science/article/pii/S0167739X21003927}
\BIBentrySTDinterwordspacing

\bibitem{mounesan2024}
\BIBentryALTinterwordspacing
M.~Mounesan, M.~Lemus, H.~Yeddulapalli, P.~Calyam, and S.~Debroy,
  ``Reinforcement learning-driven data-intensive workflow scheduling for
  volunteer edge-cloud,'' 2024. [Online]. Available:
  \url{https://arxiv.org/abs/2407.01428}
\BIBentrySTDinterwordspacing

\bibitem{survey-ml-scheduling-sla}
S.~Goodarzy, M.~Nazari, R.~Han, E.~Keller, and E.~Rozner, ``Resource management
  in cloud computing using machine learning: A survey,'' in \emph{2020 19th
  IEEE International Conference on Machine Learning and Applications (ICMLA)},
  2020, pp. 811--816.

\bibitem{reliability-microblogging}
C.~Bayliss, J.~Panadero, L.~Calvet, and J.~Marquès, ``Reliability in volunteer
  computing micro-blogging services,'' \emph{Future Generation Computer
  Systems}, vol. 115, 09 2020.

\bibitem{oneedge}
\BIBentryALTinterwordspacing
E.~Saurez, H.~Gupta, A.~Daglis, and U.~Ramachandran, ``Oneedge: An efficient
  control plane for geo-distributed infrastructures,'' in \emph{Proceedings of
  the ACM Symposium on Cloud Computing}, ser. SoCC '21.\hskip 1em plus 0.5em
  minus 0.4em\relax New York, NY, USA: Association for Computing Machinery,
  2021, p. 182–196. [Online]. Available:
  \url{https://doi.org/10.1145/3472883.3487008}
\BIBentrySTDinterwordspacing

\bibitem{mesos}
B.~Hindman, A.~Konwinski, M.~Zaharia, A.~Ghodsi, A.~D. Joseph, R.~Katz,
  S.~Shenker, and I.~Stoica, ``Mesos: a platform for fine-grained resource
  sharing in the data center,'' in \emph{Proceedings of the 8th USENIX
  Conference on Networked Systems Design and Implementation}, ser.
  NSDI'11.\hskip 1em plus 0.5em minus 0.4em\relax USA: USENIX Association,
  2011, p. 295–308.

\bibitem{application-aware-ts}
L.~Lin, P.~Li, J.~Xiong, and M.~Lin, ``Distributed and application-aware task
  scheduling in edge-clouds,'' in \emph{2018 14th International Conference on
  Mobile Ad-Hoc and Sensor Networks (MSN)}, 2018, pp. 165--170.

\bibitem{semi-markov-reliability}
T.~M. Mengistu, D.~Che, A.~Alahmadi, and S.~Lu, ``Semi-markov process based
  reliability and availability prediction for volunteer cloud systems,'' in
  \emph{2018 IEEE 11th International Conference on Cloud Computing (CLOUD)},
  2018, pp. 359--366.

\bibitem{8457994}
Y.~Alsenani, G.~V. Crosby, T.~Velasco, and A.~Alahmadi, ``Remot reputation and
  resource-based model to estimate the reliability of the host machines in
  volunteer cloud environment,'' in \emph{2018 IEEE 6th International
  Conference on Future Internet of Things and Cloud (FiCloud)}, 2018, pp.
  63--70.

\bibitem{8473386}
Y.~Alsenani, G.~Crosby, and T.~Velasco, ``Sara: A stochastic model to estimate
  reliability of edge resources in volunteer cloud,'' in \emph{2018 IEEE
  International Conference on Edge Computing (EDGE)}, 2018, pp. 121--124.

\bibitem{Tuli2020DynamicSF}
\BIBentryALTinterwordspacing
S.~Tuli, S.~Ilager, K.~Ramamohanarao, and R.~Buyya, ``Dynamic scheduling for
  stochastic edge-cloud computing environments using a3c learning and residual
  recurrent neural networks,'' \emph{IEEE Transactions on Mobile Computing},
  vol.~21, pp. 940--954, 2020. [Online]. Available:
  \url{https://api.semanticscholar.org/CorpusID:215752435}
\BIBentrySTDinterwordspacing

\bibitem{likas2003global}
A.~Likas, N.~Vlassis, and J.~J. Verbeek, ``The global k-means clustering
  algorithm,'' \emph{Pattern recognition}, vol.~36, no.~2, pp. 451--461, 2003.

\bibitem{VECreqchallim}
\BIBentryALTinterwordspacing
M.~{Nouman Durrani} and J.~A. Shamsi, ``Volunteer computing: requirements,
  challenges, and solutions,'' \emph{Journal of Network and Computer
  Applications}, vol.~39, pp. 369--380, 2014. [Online]. Available:
  \url{https://www.sciencedirect.com/science/article/pii/S1084804513001665}
\BIBentrySTDinterwordspacing

\bibitem{rnn-feedforward}
A.~ElSaid, S.~Benson, S.~Patwardhan, D.~Stadem, and T.~Desell, \emph{Evolving
  Recurrent Neural Networks for Time Series Data Prediction of Coal Plant
  Parameters}, 03 2019, pp. 488--503.

\bibitem{VECA}
``Veca: Reliable and confidential resource clustering for volunteer edge-cloud
  computing,'' \url{https://github.com/hemanth5555/VEC}, Accessed April 2024.

\end{thebibliography}

\end{document}